\documentclass[a4paper, 10pt, conference]{ieeeconf}      % Use this line for a4
\IEEEoverridecommandlockouts                              % This command is only
\usepackage{booktabs}
\usepackage{graphicx}
\usepackage{mathtools}
\usepackage{amsmath}
\usepackage{environ}
\NewEnviron{myequation}{%
    \begin{equation}
    \scalebox{1.3}{$\BODY$}
    \end{equation}
    }

\title{
\noindent\rule{14.6cm}{3.0pt}\\
\LARGE \bf
Real-time Automated Answer Scoring\\
\noindent\rule{14.6cm}{0.4pt}\\
}

\vspace{2 in}
\author{ \parbox{2.2 in}{\centering Akash Nagaraj\\
         Department of Computer Science\\
         PES University\\
         Bangalore, India\\
         {\tt akashn1897@gmail.com}}
         \parbox{2.2 in}{ \centering Mukund Sood\\
         Department of Computer Science \\
         PES University\\
         Bangalore, India\\
         {\tt mukundsood2013@gmail.com}}
          \parbox{2.2 in}{ \centering Dr. Gowri Srinivasa\\
         Department of Computer Science \\
         PES University\\
         Bangalore, India\\
         {\tt gsrinivasa@pes.edu}}
}

\begin{document}

\maketitle
\thispagestyle{empty}
\pagestyle{empty}

%%%%%%%%%%%%%%%%%%%%%%%%%%%%%%%%%%%%%%%%%%%%%%%%%%%%%%%%%%%%%%%%%%%%%%%%%%%%%%%%
\begin{abstract}
In recent years, the role of big data analytics has exponentially grown, and is now slowly making its way into the education industry. Several attempts are being made in this sphere in order to improve the quality of education being provided to students while simultaneously improving teaching techniques to reduce the burden put on the teacher. While the implications of such a collaboration have in fact been carried out before, automated scoring of answers has been explored to a rather limited extent. 

The biggest obstacle to choosing constructed-response assessments over the traditional multiple-choice assessments is the large cost, and effort that comes with their scoring, and this is precisely the issue that this project aims to solve.The aim is to accept raw-input from the student in the form of their answer, pre-process these answers if required, and automatically score the answers. Evaluation of these answers will be based purely on the previous examples on which the model has been trained. In addition, we have made this a real-time system that captures "snapshots" of the writer's progress with respect to the answer, allowing us to see how the student has arrived at his/her final answer. This additional information can allow us to analyse where students err, allowing us to better understand the thought-process of students as they take tests and possibly unearth previously unknown trends.

\end{abstract}

%%%%%%%%%%%%%%%%%%%%%%%%%%%%%%%%%%%%%%%%%%%%%%%%%%%%%%%%%%%%%%%%%%%%%%%%%%%%%%%%

\section{INTRODUCTION}

Automated answer scoring is the use of specialized computer programs to assign grades to answers written in an educational setting. It is a method of educational assessment and an application of natural language processing. \\
Attempts to build an automated grading system dated back to 1966 when Ellis B. Page proved on The Phi Delta Kappan that a computer could do as well as a single human judge [1]. Since then, much effort has been put into building the perfect system. Intelligent Essay Assessor (IEA), developed by Peter Foltz and Thomas Landauer, was first used to score essays for large undergraduate courses in 1994 [2]. The automated reader developed by the Educational Testing Service, e-Rater, used hundreds of manually defined features. It was trained on 64 different prompts and more than 25,000 answers on a 6-point scale from 1 to 6. Evaluated on the quadratic weighted kappa calculated between the automated scores for the answers and the resolved score for human raters on each set of essays, e-rater could only achieve a kappa score below 0.5. [3]\\
In 2012, the Hewlett Foundation sponsored a competition on Kaggle called the Automated Student Assessment Prize (ASAP). The competition also used quadratic weighted kappa to measure the similarity between the human scores and the automated scores. 154 participants attempted to predict the essay score. The winning team got a kappa score of 0.81407. [4] Later, a team at Carnegie Mellon University built a model using dense and sparse features, trained on the same dataset to achieve the kappa score of 0.833. [5]\\
Our approach will be different from the previous attempts to solve the problem, using deep learning, as opposed to domain oriented predefined features. The best way to improve one's own writing skill is to write an answer, receive feedback from your instructor, and based on the feedback, revise your answer. Repetition of this process as often as possible is advised up until an optimal score is reached. However, the problem arises in the fact that continuously evaluating these essays requires a lot of time and puts an enormous load on the classroom teacher who is often found in a position where she must provide feedback to a possible 30 essays or more each time a topic is assigned. As a result, teachers are not able to give writing assignments as often as they would wish. Keeping this in mind, researchers have sought to develop applications that automate essay scoring and evaluation, with the goal of allowing teachers more flexibility in designing a course so as to maximize the learning of the student without having to compromise due to a large work load. Automated scoring of answers is not meant to serve as a replacement to a human evaluator, but, merely as an aid to the evaluator.

\section{RELATED WORK}

\subsection{Recurrent Neural Networks for Language Understanding [6]}
Shows the effectiveness of Recurrent Neural Networks for language understanding. The core of their project was to take words as inputs in a standard RNN-LM and to predict slot labels instead of the traditional method of words on the output side. The RNN-LM represents each word as a high-dimensional real-valued vector in such a way that similar words tend to be closer together and relationships between words are preserved. They demonstrated the use of the Elman Architecture and using the ATIS Dataset, they were able to show that the RNN outperformed the Conditional Random Fields (CRF) approaches by a large margin as well as previous neural-net approaches. 

\subsection{Automated Essay Scoring Using Bayes' Theorem [7]}
Bayesian Models for text classification from the information science field were extended and applied to student produced essays, with each model calibrated to 462 essays with no score points. They manipulated some variables, such as trimming, stemming and the use of stopwords to imporve their accuracy. They were able to achieve an accuracy of only 80\%. The major drawbacks of this paper are that it uses a very simple Bayesian Model for the classification. 

\subsection{Cosine similarity to determine similarity measure: Study case in online essay assessment [8]}
Implementation of the weighting of Term Frequency - Inverse Document Frequency (TF-IDF) method and Cosine Similarity with the measuring degree concept of similarity terms in a document. Tests have been carried out on a number of Indonesian text-based documents that have gone through the stage of pre-processing for data extraction purposes. This process results is in a ranking of the document weight that have closeness match level with expert's document.

\section{PROBLEM STATEMENT}

The aim is to implement a  system that evaluates a student's answer $A$ and recommends a score $s(A)$, accurately as well as efficiently, such that; rather than just evaluating the end-product $A_f$, consider a sequence of intermediate answers $A_1$, $A_2$, $...$, $A_n$ (where $A_n$ = $A_f$) captured at suitable times $t_1$, $t_2$, $...$, $t_n$ as the student builds his/her answer. \\By relating $s(A_i)$ vs $t_i$ and plotting the graph, we will be able to gain insight on the thought process and steps taken by which the student arrived at his/her answer. For instance, two students may end up with the same score but one of the students may have gradually built up to this score, whereas the other may have actually written an answer that earned a higher score before making modifications that lowered the score.
\subsection{The Dataset}
The data set we are using for this project is The Hewlett Foundation: Automated Essay Scoring [4].

About the data set:
\begin{description}
\item[$\bullet$] There are 8 different sets of essays, each generated from a single prompt.
\item[$\bullet$] Selected essays range from an average length of 150 to 550 words per response.
\item[$\bullet$] Each essay was hand-graded a human evaluator, which means we are using real-world data to produce real-world results.
\item[$\bullet$] There are 10686 samples in total. We used a 85:15 data split for training and testing, as well as validation.
\item[$\bullet$] Each set has a different grading scale.
\end{description}
%%\newpage
%%\includegraphics[width=17cm, height=2.4cm]{Block_Diagram.png}
%%\hspace*{1cm} Block Diagram of the Proposed System

\subsection{Evaluation Metric [16]}In recent years, the role of big data analytics has exponentially grown, and is now slowly making its way into the education industry. There is a growing need to tap into the massive reserves of education data emanating from both private as well as government institutions. Several attempts are being made in this sphere in order to improve the quality of education being provided to students while simultaneously improving teaching techniques to reduce the burden put on the teacher. While the implications of such collaboration have in fact been carried out before, automated scoring of answers has been explored to a rather limited extent. 

The biggest obstacle to choosing constructed-response assessments over the traditional multiple-choice assessments is the large cost, and effort that comes with their scoring, and this is precisely the issue that this project aims to solve.The project is to accept raw-input from the student in the form of their answer, pre-process these answers if required, and automatically score the answers. Evaluation of these answers will be based purely on the previous examples on which the model has been trained. In addition, we have made this a real-time system that captures "snapshots" of the writer's progress with respect to the answer, allowing us to see how the student has arrived at his/her final answer. This additional information can allow us to analyse where students err, allowing us to better understand the thought-process of students as they take tests and possibly unearth previously unknown trends. 
\begin{center}
\begin{myequation}
\kappa = 1 - \frac{\sum_{ij}W_{ij}O_{ij}}{\sum_{ij}W_{ij}E_{ij}}
\end{myequation}

\end{center}
\begin{description}
\item[$\bullet$] The Weight matrix $W$ is calculated based on the difference between raters scores.\\
\item[$\bullet$] $O_{ij}$ corresponds to the number of answers that received a rating $i$ by Rater A and Rating $j$ by Rater B.\\
\item[$\bullet$] $E$ is the $n x n$ histogram matrix of expected ratings.
\end{description}
The evaluation metric used for the project is quadratic weighted kappa  [10]. The Quadratic Weighted Kappa (QWK) metric typically varies from 0 (only random agreement between raters) to 1 (complete agreement between raters), and doesn't penalize for an incorrect score match, but rather, it takes into account the error in the scores, with respect to the range of scores. Therefore, it serves as a much better evaluation metric as compared to accuracy, since in our problem, we're primarily comparing inter-rater agreement.

\section{APPROACH}
The primary objective is to build a model that accepts an answer as input and automatically outputs the score of that answer, in an efficient manner that can be done in \textit{real-time} as the student writes the answer. The score would be evaluated at the end of sentences, words or ever a few systems that only perform the analysis and evaluate the final draft in a answer.\\
Coming to the pre-processing don. What we first did, was try to build a Bag of Words Model and examine the results it would give us, but they were not very promising. Based on our findings, we decided to move forward and remove all stop-words from each of the essays being passed to our Model. We also implemented a rudimentary Spell Check on the essays based on the Peter Norvig Spell Check Algorithm and stemming, however, the results we achieved were substandard possibly owing to the fact that human evaluators would consider a misspelled word as an important factor (reducing overall marks of the answer) and we decided against using them continuing forward with our research. The interesting bit of the pre-processing comes, where we decided to implement a word-embedding model. As mentioned above we had two possible choices, GloVe [12] and Word2Vec [16]. Based on some research into the capabilities of both models we found that the Word2Vec would possibly be a better model, however, just to confirm for ourselves, we decide to implement both, working however, on the assumption that the Word2Vec Model would outperform the GolVe Model.\\
The role that Word2Vec plays in our implementation is a massive one. Developed by team of researchers at Google, it is an extremely useful group of related models to produce word vectors of your input. The Word2Vec Model is trained on the input specified, and is trained in such a way that words that have a similar meaning have a smaller vector distance between them. This representation of words is very helpful going forward with our predictions primarily because we are able to train our Model based on the type of input, rather than predefined meanings of the word, which was consequently one of the reasons that we chose this model over GloVe, thereby ensuring that words that have a similar meaning in that context or tend to appear together frequently have smaller distances than words that are similar by their intrinsic meaning.\\

\begin{table}
\centering
\caption{Random Forest v/s SVM RBF}
\label{my-label}
\begin{tabular}{@{}lll@{}}
\toprule
                          			& \textbf{Word2Vec} 	&            						\\ \midrule
\textbf{Model}                    	& \textbf{QWK}      	& \textbf{Dim} 	\\
\textbf{Random Forests}           	& 0.9535   				& 300        			\\
\textbf{Random Forests}           	& 0.9563   				& 200        			\\
\textbf{Random Forests}            	& 0.9508   				& 100        			\\
                          			&          				&            			\\
\textbf{SVM Radial Basis Function} 	& 0.9619   				& 300        			\\
\textbf{SVM Radial Basis Function} 	& 0.9568   				& 200        			\\
\textbf{SVM Radial Basis Function} 	& 0.9509   				& 100        			\\ \bottomrule
\end{tabular}
\end{table}

Once we obtained the vector representations of our input, we used Vector Averaging to get a specific Vector Value for each essay in the dataset, which we could then use to make better predictions. In our approach we used four Models; a Random Forest Classifier, a Radial Basis Support Vector Machine, a Deep Neural Network and a Short Long Term Memory Unit which is a form of a Recurrent Neural Network. 
The Random Forest Classifier [17] and the Support Vector Machine [18] returned very good results. Looking into Random Forest Classifiers, they are an example of an ensemble learner built on decision trees, is a direct consequence of the fact that by maximum voting from a panel of independent judges, we get the final prediction better than the best judge.\\
Our Deep Neural Network is a simple two-layered feed forward neural network, where we use the 'Rectified Linear Units' as the activation function for the first layer and a 'Softmax' activation function [19] for the output layer to squash the output vector. We tried multiple activation functions, but verified that these activation functions give us an optimum result and the loss function that we have used is 'sparse categorical crossentropy' since our problem deals with more than two classes. While building the Word2Vec Model we achieved the best results when the dimension for each word vector was set to 300. Once trained, the Neural Network architecture would be used to evaluate 'snapshots' of the students' intermediate responses which will give us a graph of the student's score vs time intervals, which could answer many interesting questions with respect to the student's performance.\\
\begin{table}
\centering
\caption{LSTM v/s DNN v/s Combined Model}
\label{lstm_dnn}
\begin{tabular}{@{}llllll@{}}
\toprule
                       		& \textbf{Word2Vec} &\textbf{ GloVe}    &        &            							\\ \midrule
\textbf{Model}                  & \textbf{QWK}    & \textbf{QWK}    & \textbf{Dim} 		\\
\textbf{Long Short Term Memory} & 0.9653   & 0.9566 & 300       											\\
\textbf{Long Short Term Memory} & 0.9642   & 0.9623 & 200        											\\
\textbf{Long Short Term Memory} & 0.9512   & 0.9433 & 100        											\\
                       			&          &        &            											\\
\textbf{Deep Neural Network}    & 0.9643   & 0.9555 & 300        											\\
\textbf{Deep Neural Network }   & 0.9576  & 0.955  & 200        											\\
\textbf{Deep Neural Network}    & 0.9611  & 0.9442 & 100        											\\
                       			&    	  &        &            											\\
\textbf{Combined Model}    		& 0.9721 & - 	& 100        											\\ \bottomrule
\end{tabular}
\end{table}
Recurrent networks take as their input not just the current input example they see, but also what they have perceived previously in time. The decision a recurrent net reached at time step t-1 affects the decision it will reach one moment later at time step t. Hence, RNNs have two sources of input, the present and the recent past, which combine to determine how they respond to new data. This quality of an RNN is what allows it to produce robust models when it comes to textual input, because of its sequential quality. The sequential information is preserved in the RNN's hidden state, which manages to span many time steps as it cascades forward to affect the processing of each new example. It is finding correlations between events separated by many moments, and these correlations are called “long-term dependencies”, because an event downstream in time depends upon, and is a function of, one or more events that came before.\\
In the mid-90s, a variation of recurrent net, called Long Short Term Memory units (LSTM) was proposed as a solution to the vanishing gradient problem [24]. LSTMs help preserve the error that can be back-propagated through time and layers. By maintaining a more constant error, they allow RNNs to continue to learn over many time steps (over 1000), thereby opening a channel to link causes and effects remotely. This feature of an LSTM, aids the RNN in the model we have built to allow for better predictions of each answer. \\
Now, considering that we are trying to predict a grade for a student, a simple accuracy test where we compare the number of correct predictions of our model, would not be a very good metric, therefore, as specified in the source of the dataset, have used Quadratic Weighted Kappa as our Evaluation Metric. Quadratic Weighted Kappa allows us to calculate the agreement between evaluator, which in our case would be the agreement of the score predicted by our models, against the score predicted by an evaluator whose scores we are using to train the model. This turns out to be a model far more similar to a real-world scenario where agreement between raters is considered, taking into account the weighted errors.\\
Finally, we used a method similar to Weighted Average. We obtained the predictions from each Model that we built, namely the LSTM Mode, the Deep Neural Network Mode, the SVM Radial Basis Function Model and the Random Forrest Classifier Model, along with their respective Quadratic Weighed Kappa values, and applying the formula:
\begin{center}
\begin{myequation}
\kappa_{f} = \frac{\kappa_{1}*S_{1} + \kappa_{2}*S_{2} + \kappa_{3}*S_{3} + \kappa_{4}*S_{4}}{\sum_{i=1}^{4} \kappa_{i}}
\end{myequation}

\end{center}
we were able to obtain better predictions for each essay that resulted in an overall higher Quadratic Weighted Kappa Score.

\subsection{General Classifier Results}
Table 2 above shows us the QWK of the Random Forrest Classifier and the SVM using a Radial Basis Function. The high QWK was unexpected, however we attribute it towards the pre-processing done as well as the use of Word Vectors. 

\section{EXPERIMENTS AND RESULTS}
Based on the above approaches, we ran multiple tests to figure out which Model would give the best results. We documented all the results and we now provide justifications for each of them, along with a couple other pre-processing techniques that we used, and how they fared as well. Earlier, we had justified our selection of Word2Vec over the pre-trained GloVe Vectors, but as a confirmation, we ran tests on both models just to take a look at how they performed under the same conditions.

\subsection{Effect of Pre-processing}
We used a couple of different pre-processing techniques which affected our results, so we feel it is appropriate to discuss these before our final Results
\subsubsection{Stemming [20]}
Stemming is the process of reducing inflected words to their word stem, base or root form. When we applied this method we incurred a drop in the QWK across all models. The reason behind this drop lies in the importance of the way in which an answer is written by a student. A human scorer would evaluate students who have better grammar and use of words higher than those who did not. Hence, when we tested the model after stemming the training and testing sets, we incurred a considerable drop in accuracy.
\subsubsection{Spell-Correction [11]}
To implement Spell-Correction we used Peter Norvig's Spell Check Algorithm [11], and this method as well, similarly to Stemming, resulted in a drop of accuracy. The reason for this is identical to the one listed above, and hence, we deiced not to use either of these methods moving forward with our testing.

\begin{figure}
\includegraphics[width=8cm, height=5cm]{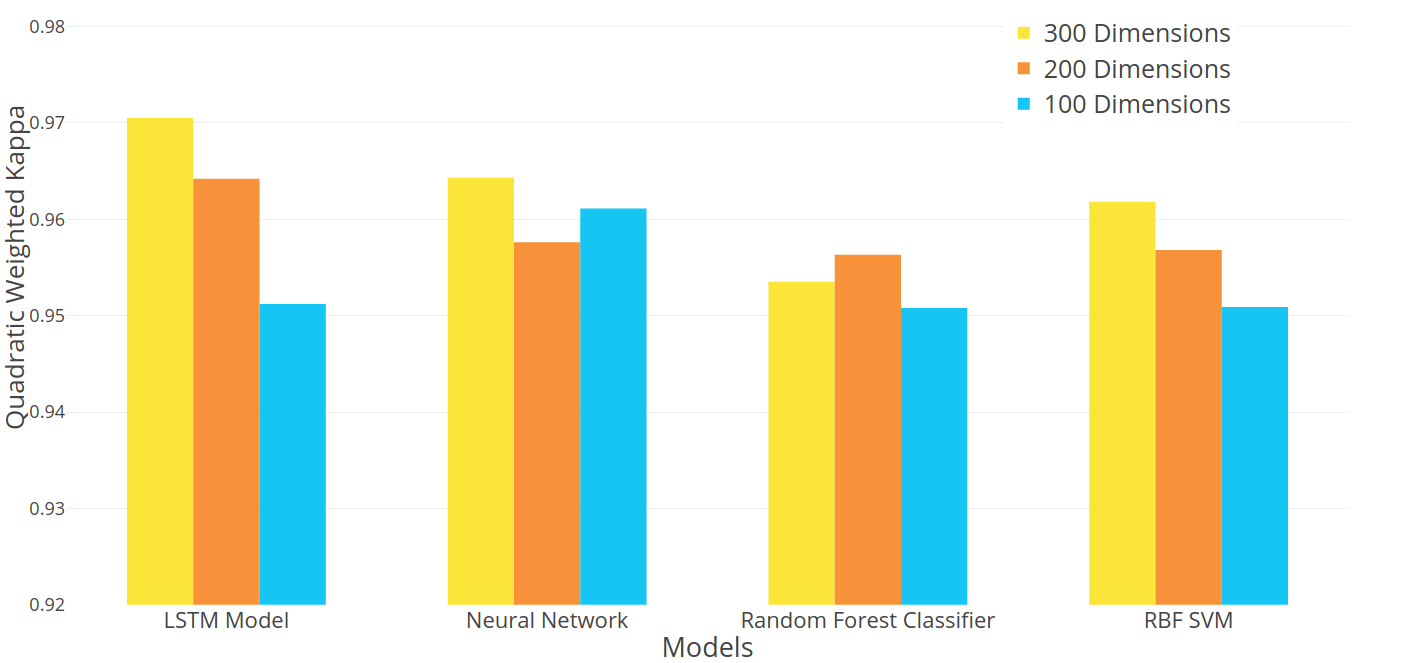}
\caption{Comparison of the different models}
\end{figure}

\subsection{Neural Network Results}
Table 1 above, has a comparison of how the LSTM Model and Deep Neural Network Model worked with respect to both Word Embedding Models. Based on the results, the following two observations can be made:
\subsubsection{Word Embedding Model}
You can see that in general the Word2Vec Model outperforms the GloVe Model. This is because Word2Vec was trained on our own dataset whereas the GloVe Model used pre-trained Word Vectors. This phenomenon is best explained with an example. Let's say we have two words, \textit{threads} and \textit{string}, in terms of Computers (which was the topic of most of the answers in our dataset) they have a completely different meaning to what they do in day-to-day life. Hence, while predicting a score, the Word2Vec Model allows us to use the relevant word in the relevant situation which GloVe does not, leading to a slightly lower Quadratic Weighted Kappa.
\subsubsection{Prediction Model}
We see that the LSTM outperforms the DNN, this increase in QWK can be attributed towards the fact that an LSTM is basically a Recurrent Neural Network, this means that it works best with sequential data, and in our case, text is a perfect example of sequential data. Additionally, an LSTM has a memory unit which allows it to take into account a larger context, thereby giving slightly better results. However we see that the Combined Model yields the best overall results, due to the fact that it takes into account the prediction made by each Model and gives a weighted average that tends to be the best fit for the Answer in question.
\begin{figure}
\includegraphics[width=8cm, height=6cm]{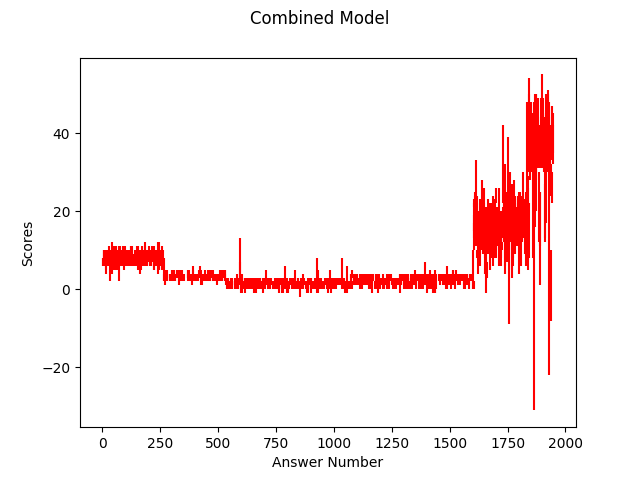}
\caption{Errors of the Combined Model using Word2Vec}
\end{figure}

\subsection{Graphs}
The visualizations seen is a general error graph of the predictions obtained using the Combined Model.\\
Finally you can see our results compared to the previous best results of the competitors who participated in Kaggle's competition. \\
We also added a graph of QWK v/s the Model used which is a comparison of all our Models compared to previous results. As illustrated in the graph, our Model has exceeded previous prediction Model by a large margin.

\subsection{Optimizers}
With regard to optimizers, we used a numerous predefined optimizers such as $Stochastic Gradient Descent$ [21], $Adam$ [22], $Adagrad$, $AdaDelta$ [23] etc. But $RMSProp$ yielded the best results, with a Quadratic Weighted Kappa of about 0.974

\subsection{Real-Time System}
We implemented our Real-Time system using a web-server designed on python's flask web framework. We grade the scores of students as they type at intervals of 0.66 seconds based on studies that show this to be the typing speed per word of the average human. The Real-time system predicts the score of "snapshots" of a student's answer using the combined model in about 0.3 seconds (running on a Intel i7-7820 Processor), making it an interactive system.

\section{Conclusion}
The inspiration behind this project was to make the job of a teacher easier. The crux of a formal education is written assignments, and teachers often are not able to give as many as they would like. Using a system built around Automated Scoring, we could move into an era where teachers can Model courses however they wish without being restricted by their workload.\\
We also find that this project could be helpful in getting to know how a student answers particular questions, and in what way they first approach a particular problem. This would give us a huge amount of data with respect to how students think, this giving us a better idea on how to teach certain concepts to students, in ways that they best understand.

\addtolength{\textheight}{-12cm}   % This command serves to balance the column lengths

%%%%%%%%%%%%%%%%%%%%%%%%%%%%%%%%%%%%%%%%%%%%%%%%%%%%%%%%%%%%%%%%%%%%%%%%%%%%%%%%

%%%%%%%%%%%%%%%%%%%%%%%%%%%%%%%%%%%%%%%%%%%%%%%%%%%%%%%%%%%%%%%%%%%%%%%%%%%%%%%%

%%%%%%%%%%%%%%%%%%%%%%%%%%%%%%%%%%%%%%%%%%%%%%%%%%%%%%%%%%%%%%%%%%%%%%%%%%%%%%%%

\end{document}